\documentclass[twocolumn,amsmath,amssymb,]{revtex4}

\usepackage{graphicx}
\usepackage{dcolumn}
\usepackage{bm}

\begin{document}

\preprint{}

\title{Quintessence reconstruction of the new agegraphic dark energy model}

\author{Jian-Pin Wu}
 \email{jianpinwu@yahoo.com.cn}
\author{Da-Zhu Ma}
 \email{mdzhbmy@126.com}
\author{Yi Ling}%
 \email{yling@ncu.edu.cn}
\affiliation{ Center for  Gravity and Relativistic Astrophysics,
Department of Physics, Nanchang University, Nanchang 330031, China
}%

\begin{abstract}
In this paper we implement the new agegraphic dark energy model with
quintessence field. We demonstrate that the new agegraphic evolution
of the universe can be described completely by a single quintessence
field. Its potential as a function of the quintessence field is
reconstructed numerically. In particular, the analytical solution of
the new agegraphic quintessence dark energy model (NAQDE) is
approximately obtained in the matter-dominated epoch. Furthermore,
we investigate the evolution of the NAQDE model in the
$\omega-\omega'$ phase plane. It turns out that by quantum
corrections, the trajectory of this model lies outside the thawing
and freezing regions at early times. But at late times, it enters
the freezing regions and gradually approaches to a static
cosmological constant state in the future. Therefore the NAQDE
should belong to the freezing model at late times. For comparison,
we further extend this model by including the interaction between
the NADE and DM and discuss its evolution in the $\omega-\omega'$
phase plane.
\end{abstract}

\pacs{Valid PACS appear here} 
\maketitle

\section{Introduction}

Recent astronomical observations indicate that the universe is
undergoing accelerated expansion at the present time \cite{1}.
Nowadays it is the most accepted idea that a mysterious dominant
component, dark energy (DE) with negative pressure, leads to this
cosmic acceleration. However, the nature and cosmological origin of
DE still remain enigmatic at present. It is not clear yet whether DE
can be described by a cosmological constant which is independent of
time, or by dynamical scalar fields such as quintessence, K-essence,
tachyon, phantom, ghost condensate or quintom. At the same time, the
so-called fine-tuning problem and coincidence problem still confuse
us.

To shed light on these fundamental and difficult problems, some
interesting DE models were proposed recently, including holographic
dark energy model \cite{2} and agegraphic dark energy model
\cite{3,4}. The former is motivated from the holographic hypothesis.
The later is constructed in the light of the
K$\acute{a}$rolyh$\acute{a}$zy relation \cite{5} and a corresponding
energy fluctuations of space-time. Therefore, although a complete
theory of quantum gravity is not established yet today, we still can
make some attempts to investigate the nature of dark energy
according to some principles of quantum gravity. The holographic
dark energy model and the agegraphic model are just such examples,
which are originated from some considerations of the features of the
quantum theory of gravity. That is to say, the holographic and
agegraphic dark energy model possess some significant features of
quantum gravity.

On the other hand, as we all admitted, the scalar field model is an
effective description of an underlying theory of dark energy. Scalar
fields naturally arise in particle physics including supersymmetric
field theories and string/M theory. Therefore, scalar field is
expected to reveal the dynamical mechanism and the nature of dark
energy. However, although fundamental theories such as string/M
theory do provide a number of possible candidates for scalar fields,
they do not uniquely predict its potential $V(\phi)$. Therefore it
becomes meaningful to reconstruct $V(\phi)$ from some dark energy
models possessing some significant features of the quantum gravity
theory, such as holographic and agegraphic dark energy model.

Some works have been investigated on the reconstruction of the
potential $V(\phi)$ in holographic dark energy models. For instance
we refer to \cite{6,7}. In this paper, we intend to reconstruct the
potential of quintessence $V(\phi)$ for agegraphic dark energy
models.

Let us first start with a close look at the agegraphic dark energy
model. According to the K$\acute{a}$rolyh$\acute{a}$zy relation,
the distance $t$ in Minkowski spacetime cannot be known to a
better accuracy than
\begin{equation}
\delta t = \lambda t ^{2/3} _{p} t ^{1/3}, \label{eq1}
\end{equation}
where $\lambda$ is a dimensionless constant of order unity. We use
the units $\hbar = c = k_{B} = 1$ throughout this work. Thus $l_{p}
= t_{p} = 1/M_{p}$ with $l_{p}$ , $t_{p}$ and $M_{p}$ being the
reduced Planck length, time and mass respectively.

Following \cite{8,9}, corresponding to Eq.(\ref{eq1}) a length scale
$t$ can be known with a maximum precision $\delta t$ determining
thereby a minimal detectable cell $\delta t^{3} \sim t^{2}_{p} t$
over a spatial region $t^{3}$. Such a cell represents a minimal
detectable unit of spacetime over a given length scale $t$. If the
age of the Minkowski spacetime is $t$, then over a spatial region
with linear size $t$ (determining the maximal observable patch)
there exists a minimal cell $t^{3}$, the energy of which due to
time-energy uncertainty relation can not be smaller than
\begin{equation}
E_{\delta t^{3}} \sim t^{-1}.\label{eq2}
\end{equation}

Therefore, the energy density of metric fluctuations of Minkowski
spacetime is given by
\begin{equation}
\rho_{q} \sim \frac{E_{\delta t^{3}}}{\delta t^{3}} \sim
\frac{1}{t_{p}^{2} t^{2}} \sim \frac{M_{p}^{2}}{t^{2}},\label{eq3}
\end{equation}
which for $t_{0}\sim H_{0}^{-1}$ gives pretty good value for the
present dark energy. We refer to the original papers \cite{8,9} for
more details.

Here some arguments are given on cosmological implications of the
K$\acute{a}$rolyh$\acute{a}$zy uncertainty relation (\ref{eq1}) and
the energy density (\ref{eq3}). First, the
K$\acute{a}$rolyh$\acute{a}$zy relation (\ref{eq1}) obeys the
holographic black hole entropy bound \cite{9}: the relation
(\ref{eq1}) gives a relation between $\delta l$ (UV cutoff) and the
length scale $l$ (IR cutoff) of a system, $\delta l \sim
l_{p}^{2/3}l^{1/3}$; the system has entropy
\begin{equation}
S \leq (\frac{l}{\delta l})^{3}\sim (\frac{l}{l_{p}})^{2}\sim
S_{BH},\label{eq38}
\end{equation}
which is less than the black hole entropy with horizon radius $l$.
Therefore, the K$\acute{a}$rolyh$\acute{a}$zy uncertainty relation
(\ref{eq1}) is a reflection of interplay between UV scale and IR
scale in effective quantum field theory \cite{10}. The microscopic
energy scales of quantum mechanics and the macroscopic properties of
our present universe are intimately connected. Second, the energy
density (\ref{eq3}) is dynamically tied to the large scales of the
universe, thus violating naive decoupling between UV scale and IR
scale. The appearance of both the Planck length and the largest
observable scale in the energy density (\ref{eq3}) seems to suggest
that the dark energy is due to an entanglement between ultraviolet
and infrared physics \cite{11}. Therefore, we expect that the
interplay between UV scale and IR scale can give us some clues about
what is the reason that quantum gravity effects are still valid
today at large distance scales. Some expectations are born out by
explicit constructions of effective field theories from string
theory \cite{10}.

Based on the energy density (3), a so-called agegraphic dark energy
(ADE) model was proposed in \cite{3}. There, as the most natural
choice, the time scale $t$ in Eq.(\ref{eq3}) is chosen to be the age
of the universe
\begin{equation}
T = \int_{0}^{a}\frac{da}{Ha},\label{eq4}
\end{equation}
where $a$ is the scale factor of our universe and H $\equiv$
$\dot{a}$/a is the Hubble parameter, where a dot denotes the
derivative with respect to cosmic time.

Later, a new model of ADE was proposed in \cite{4}, where the time
scale in Eq.(\ref{eq3}) is chosen to be the conformal time $\eta$
instead of the age of the universe. This new agegraphic dark energy
(NADE) contains some new features different from the ADE
\cite{4,12,13} and overcome some unsatisfactory points. For
instance, the ADE suffers from the difficulty to describe the
matter-dominated epoch while the NADE resolved this issue.

In this paper we intend to present a more detailed investigation on
the features of the NADE models through a quintessence
reconstruction. Our paper is organized as follows. We first present
brief review on the NADE in section $2$. Then we demonstrate a
correspondence between the NADE scenario and quintessence dark
energy model in section $3$. The potential of the new agegraphic
quintessence is reconstructed numerically. In addition, we provide
an analytical solution of the NAQDE in the matter-dominated epoch.
In section $4$, we investigate the evolution of the NAQDE model in
the $\omega-\omega'$ plane and obtain some important characteristics
on the NAQDE model. For comparison we consider the NADE with
interaction and investigate its evolution in the $\omega-\omega'$
plane in section $5$. Finally, the summary follows in section $6$.

\section{review of the new agegraphic dark energy}
In the NADE model \cite{4}, the energy density of NADE is
\begin{equation}
\rho_{q} = \frac{3n^{2}M_{p}^{2}}{\eta^{2}},\label{eq5}
\end{equation}
where the conformal time $\eta$ is given by
\begin{equation}
\eta\ \equiv \int\frac{dt}{a} = \int\frac{da}{a^{2}H}.\label{eq6}
\end{equation}
Note that $\dot{\eta} = 1/a$.

The numerical factor 3$n^{2}$ is introduced to parameterize some
uncertainties, such as the species of quantum fields in the
universe, the effect of curved spacetime (since the energy density
in Eq.(\ref{eq3}) is derived for Minkowski spacetime), and so on. In
\cite{13}, H. Wei and R. G. Cai find that the coincidence problem
can be solved naturally in the NADE model provided that $n$ is of
order unity. In addition, they constrain NADE by using the
cosmological obervations of SNIa, CMB and LSS. The joint analysis
gives the best-fit parameter (with 1$\sigma$ uncertainty)
$n=2.716_{-0.109}^{+0.111}$. And in \cite{14}, K. Y. Kim $\emph{et
al.}$ argued that the NADE model could describe the matter-dominated
 (radiation-dominated) universe in the far past only when the
parameter $n$ is chosen to be $n>n_{c}$, where the critical values
are determined to be $n_{c}=2.6878(2.5137752)$ numerically.

A flat FRW (Friedmann-Robertson-Walker) universe composed of the
NADE $\rho_{q}$ and the pressureless matter $\rho_{m}$ is governed
by the Friedmann equation
\begin{equation}
3M_{p}^{2}H^{2} = \rho_{m} + \rho_{q}.\label{eq7}
\end{equation}

Introducing the fractional energy densities $\Omega_{i} \equiv
\rho_{i}/(3M_{p}^{2}H^{2})$ for i = m and q, then one finds
\begin{equation}
\Omega_{q} = \frac{n^{2}}{H^{2}\eta^{2}}.\label{eq8}
\end{equation}

Combining Eqs.(\ref{eq5}), (\ref{eq6}), (\ref{eq7}) and (\ref{eq8}),
and using the energy conservation equation $\dot{\rho_{m}}$ +
3$H\rho_{m}$ = 0, we find that the equation of motion for
$\Omega_{q}$ is given by
\begin{equation}
\frac{d\Omega_{q}}{da} =
\frac{\Omega_{q}}{a}(1-\Omega_{q})(3-\frac{2}{n}\frac{\sqrt{\Omega_{q}}}{a}).\label{eq9}
\end{equation}

We can also rewrite Eq.(\ref{eq9}) as
\begin{equation}
\frac{d\Omega_{q}}{dz} =
-\Omega_{q}(1-\Omega_{q})[3(1+z)^{-1}-\frac{2}{n}\sqrt{\Omega_{q}}],\label{eq10}
\end{equation}
where $z = 1/a - 1$ is the redshif of the universe, and we have set
the present scale factor $a_{0} = 1$. This equation describes the
behavior of the NADE completely.

From the energy conservation equation $\dot{\rho_{q}} + 3H(\rho_{q}
+ p_{q}) = 0$, as well as Eqs.(\ref{eq5}) and (\ref{eq8}), it is
easy to find that the equation-of-state of NADE $\omega_{q} \equiv
p_{q}/\rho_{q}$ is given by
\begin{equation}
\omega_{q} = -1 +
\frac{2}{3n}\frac{\sqrt{\Omega_{q}}}{a},\label{eq11}
\end{equation}

or
\begin{equation}
\omega_{q} = -1 + \frac{2}{3n}\sqrt{\Omega_{q}}(1 + z).\label{eq12}
\end{equation}

As is referred in \cite{4}, $-1\leq\omega_{q}\leq-\frac{2}{3}$.
Thus, the NADE can be described by the quintessence. As an
illustrative example, we plot in Fig.1 the evolutions of the
equation of state (11) of the NADE with the cases $n = 3.0, 3.5,
4.0$ and $5.0$. It is clear to see that $\omega_q$ always evolves in
the region $-1\leq\omega_{q}\leq-\frac{2}{3}$, and converges to $-
1$.

\begin{figure}
\center{
\includegraphics[scale=0.8]{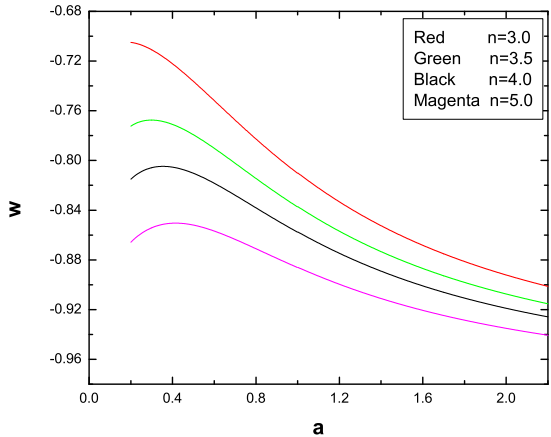}
\caption{\label{fig:wide}the evolutions of the equation of state of
the NADE. Here we take $\Omega_{q0} = 0.73$, and  show the cases $n
= 3.0, 3.5, 4.0$ and $5.0$. Clearly, the cases always evolve in the
region $-1 \leq \omega_{q} \leq - \frac{2}{3}$, and converge to $-
1$.} }
\end{figure}

\section{new agegraphic quintessence}

In this section, we link the new agegraphic dark energy model with
quintessence fields, forming a new agegraphic quintessence dark
energy model. Then we numerically simulate the evolution of the
quintessence field with explicit forms of the potential $V(\phi)$.
The analytical solution of the NAQDE in the matter-dominated epoch
is also obtained.

\subsection{Reconstructing new agegraphic quintessence dark energy}

The energy density and pressure for the quintessence field are as
following
\begin{equation}
\rho_{\phi} = \frac{1}{2}\dot{\phi}^{2} + V(\phi),\label{eq13}
\end{equation}
\begin{equation}
p_{\phi} = \frac{1}{2}\dot{\phi}^{2} - V(\phi).\label{eq14}
\end{equation}
Then one can obtain
\begin{equation}
V(\phi) = \frac{1-\omega_{\phi}}{2}\rho_{\phi},\label{eq15}
\end{equation}
\begin{equation}
\dot{\phi}^{2} = (1+\omega_{\phi})\rho_{\phi}.\label{eq16}
\end{equation}
Moreover, from the equation of FRW $3M_{p}^{2}H^{2} = \rho_{m} +
\rho_{\phi}$, we can obtain
\begin{equation}
E(a) \equiv \frac{H(a)}{H_{0}} =
(\frac{\Omega_{m0}}{(1-\Omega_{\phi})a^{3}})^{1/2},\label{eq17}
\end{equation}
where $\Omega_{m0}$ denotes the fractional energy density of
pressureless matter today.

Using Eq.(\ref{eq17}), one can rewrite Eqs.(\ref{eq15}) and
(\ref{eq16}) respectively
\begin{equation}
\frac{V(\phi)}{\rho_{c0}} =
\frac{1}{2}(1-\omega_{\phi})\Omega_{\phi}E^{2},\label{eq18}
\end{equation}
\begin{equation}
\frac{\dot{\phi}^{2}}{\rho_{c0}} =
(1+\omega_{\phi})\Omega_{\phi}E^{2},\label{eq19}
\end{equation}
where $\rho_{c0} = 3M_{P}^{2}H_{0}^{2}$ is critical density of the
universe at present.

Now we suggest a correspondence between the NADE and the
quintessence scalar field, namely, we identify $\rho_{\phi}$ with
$\rho_{q}$. Then, the quintessence field acquires the new agegraphic
nature such that $E$, $\Omega_{\phi}$ and $\omega_{\phi}$ are given
by Eqs.(\ref{eq10}), (\ref{eq12}) and (\ref{eq17}). Without loss of
generality, we assume $\dot{\phi} > 0$ in this paper \cite{6,15,16}.
Then, the derivative of the scalar field $\phi$ with respect to the
scale factor $a$ can be given by
\begin{equation}
\frac{\frac{d\phi}{da}}{M_{p}} =
\frac{\sqrt{3(1+\omega_{\phi})\Omega_{\phi}}}{a}.\label{eq20}
\end{equation}
Consequently, we can easily obtain the evolutionary form of the
field by integrating the above equation
\begin{equation}
\phi(a) = \int_{1}^{a}\frac{d\phi}{da}da,\label{eq21}
\end{equation}
where the field amplitude at the present epoch ($a=1$) is fixed to
be zero, namely $\phi(1) = 0$.

In this way, we establish a new agegraphic quintessence dark energy
model and reconstruct the potential of the NAQDE.

\subsection{The numerical solution of the NAQDE}

The quintessence models with different potential forms have been
discussed widely in the literature. For the NAQDE model constructed
in this paper, the potential V($\phi$) can be determined by Eqs.
(\ref{eq18}), (\ref{eq19}) and (\ref{eq20}), and the evolution of
$\Omega_{\phi}$ and $\omega_{\phi}$ are determined by Eqs.
(\ref{eq9}) and (\ref{eq11}) respectively. The analytical form of
the potential V$(\phi)$ is hard to be derived due to the complexity
of these equations, but we can obtain the new agegraphic
quintessence potential numerically. According to Eqs.(\ref{eq20})
and (\ref{eq21}), $\phi(a)$ is displayed in Fig.2. The reconstructed
quintessence potential $V(\phi)$ is plotted in Fig.3. Selected
curves are plotted for the cases of $n = 2.7, 3, 3.5, 4$ and $5$,
and the present fractional matter density is chosen to be
$\Omega_{mo} = 0.27$. From Fig.2 and Fig.3, we can see that the
reconstructed quintessence potential is steeper in the early epoch
and becomes very flat near today. Consequently, the scalar field
$\phi$ rolls down the potential with the kinetic energy $\dot{\phi}$
gradually decreasing. Furthermore, we can also find that the $n$ is
smaller, the potential $V(\phi)$ is more flat near today. In the
next section, we will discuss the approximate solution of the model
in the matter-dominated epoch.

\begin{figure}
\center{
\includegraphics[scale=0.8]{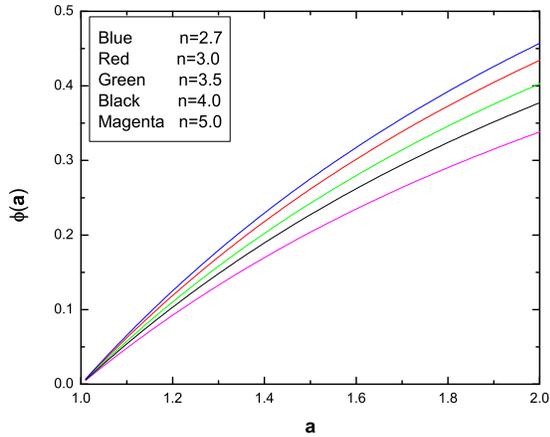}
\caption{\label{fig:wide}The revolutions of the scalar-field
$\phi(a)$ for the NAQDE, where $\phi$ is in unit of $M_{p}$. We take
here $\Omega_{mo} = 0.27$.} }
\end{figure}

\begin{figure}
\center{
\includegraphics[scale=0.8]{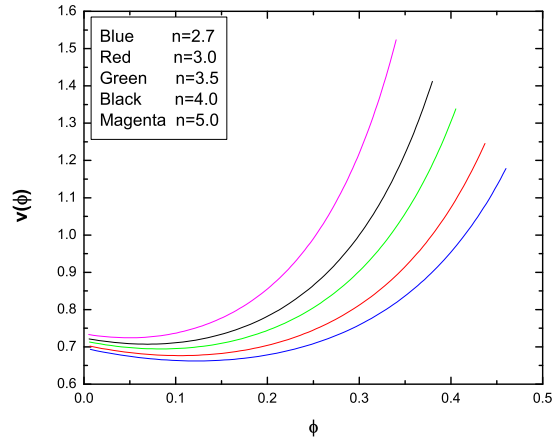}
\caption{\label{fig:wide}The reconstruction of the potential for the
NAQDE, where $\phi$ is in unit of $M_{p}$ and V$(\phi)$ in
$\rho_{c0}$. We take here $\Omega_{m0} = 0.27.$} }
\end{figure}

\subsection{Approximate solution of the NAQDE in the matter-dominated epoch}

As shown in section $4$, the quantum effect is more obvious at early
times than at late times. Therefore it is intriguing to considerate
the form of the potential $V(\phi)$ in the matter-dominated epoch.

As pointed out in \cite{4}, In matter-dominated epoch,
$\omega_{\phi}\approx-2/3$, $\Omega_{\phi}=\frac{n^{2}}{4}a^{2}$.
From Eqs.(\ref{eq20}) and (\ref{eq18}), one can obtain respectively
\begin{equation}
\phi=-\frac{nM_{p}}{2}a+\alpha,\label{eq22}
\end{equation}
\begin{equation}
V(a)=\frac{5}{24}\rho_{c0}\Omega_{m0}n^{2}\frac{a^{-3}}{a^{-2}-\frac{n^{2}}{4}},\label{eq23}
\end{equation}
where $\alpha$ is an integration constant.

Combining Eqs.(\ref{eq22}) and (\ref{eq23}), the potential $V(\phi)$
can be derived as
\begin{equation}
V(\phi)=\frac{\beta(\alpha-\phi)^{-3}}{M_{p}^{2}(\alpha-\phi)^{-2}-1},\label{eq24}
\end{equation}
where $\beta=\frac{5}{48}\rho_{c0}\Omega_{m0}n^{3}M_{p}^{3}$.

It is intriguing to see the expansion of the potential $V(\phi)$
around $\phi=\alpha$. Setting $\varphi=\alpha-\phi$, the potential
around $\varphi=0$ can be expanded as
\begin{equation}
V(\varphi)=\frac{\beta}{M_{p}^{3}}[(\frac{\varphi}{M_{p}})^{-1}+\frac{\varphi}{M_{p}}+(\frac{\varphi}{M_{p}})^{3}+(\frac{\varphi}{M_{p}})^{5}+(\frac{\varphi}{M_{p}})^{7}+\cdots].\label{eq25}
\end{equation}

This leads us to surmise that, in reality, $V(\phi)$ around
$\phi=\alpha$ might be considered as an effective potential
resulting from the combination of some different fields.

From Eq.(\ref{eq25}), we can see that the first term is tracking
potential and the other terms are not tracking potential when
$\omega_{\phi}<\omega_{m}$\cite{17}. This implies that the potential
of the NAQDE in the matter-dominated epoch is not tracking
potential. In fact, we can easily find that $\Gamma(\phi)$ does not
always satisfy $\Gamma(\phi)>1$, where
$\Gamma(\phi)=\frac{V(\phi)''V(\phi)}{(V(\phi)')^{2}}$. Therefore
the potential of the NAQDE in the matter-dominated epoch is not a
tracking solution\cite{17}.

\begin{figure*}[tbph]
\center{
\includegraphics[scale=0.8]{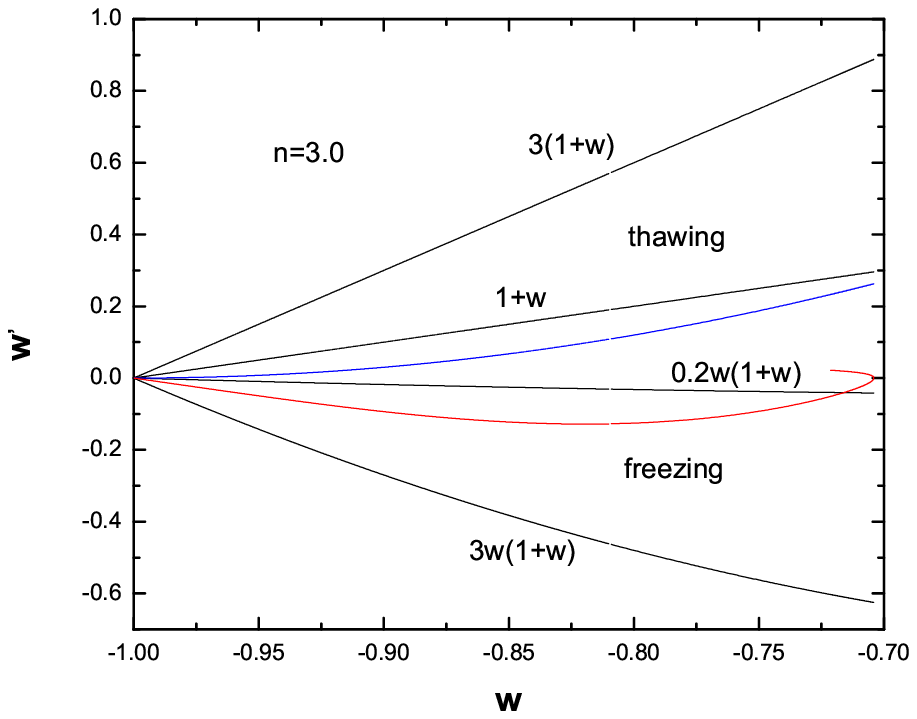}
\includegraphics[scale=0.8]{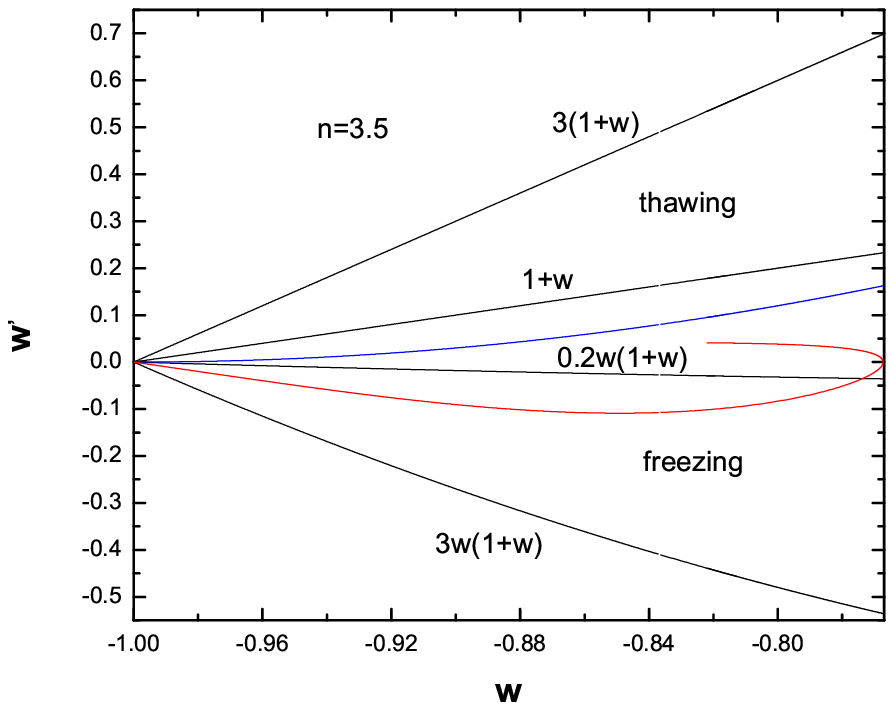}
\includegraphics[scale=0.8]{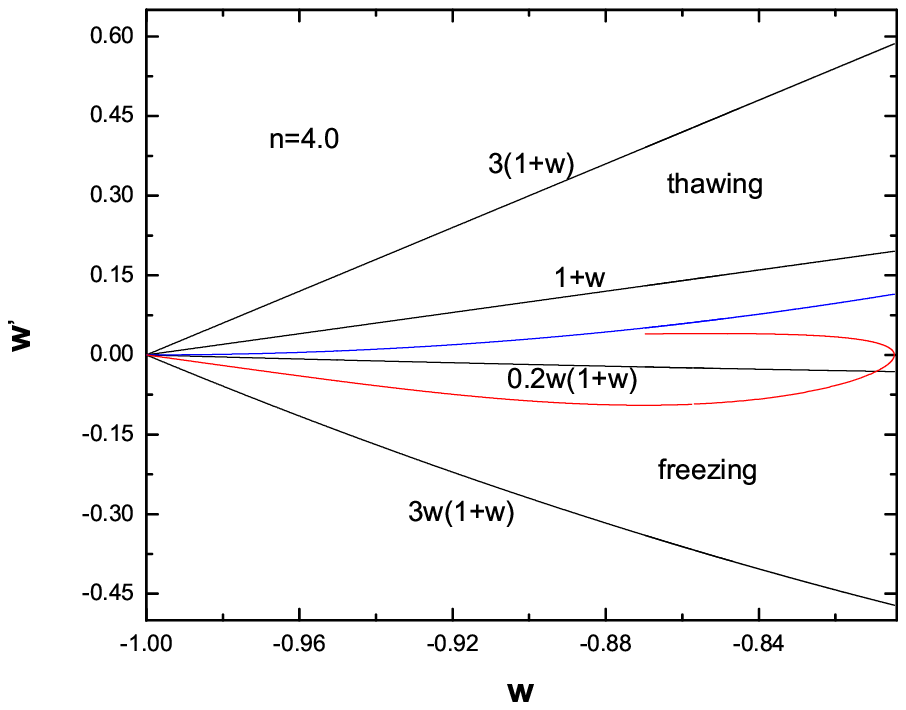}
\includegraphics[scale=0.8]{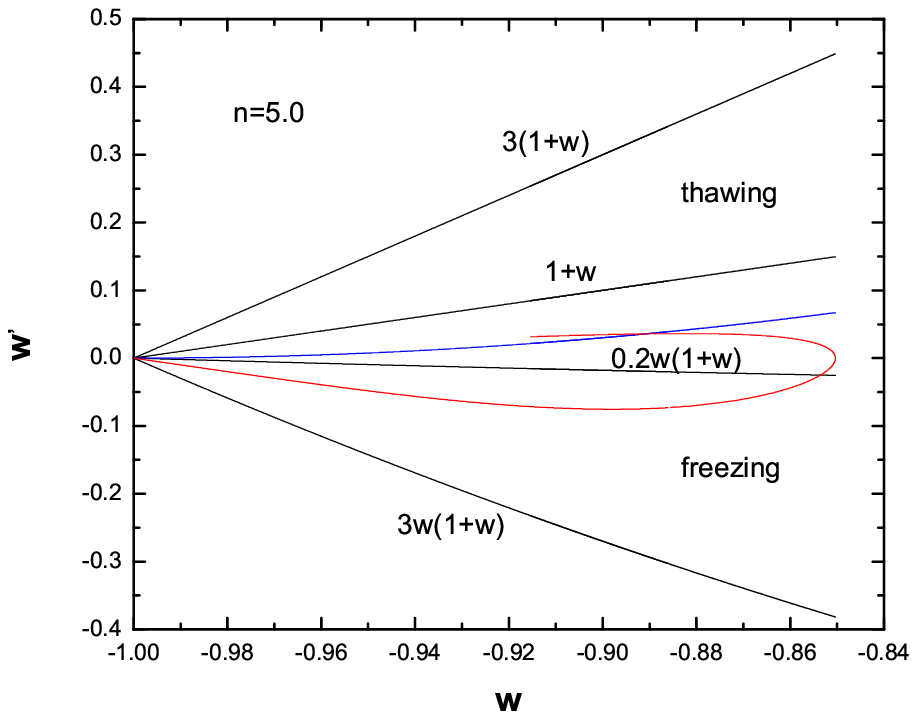}
\caption{The trajectory of the NAQDE in $\omega - \omega'$ plane,
where selected curves are plotted for the cases of $n = 3.0, 3.5,
4.0$ and $5.0$, respectively. Here we take $\Omega_{\phi0}=0.73$.
 } }
\end{figure*}

\section{the evolution of the NAQDE in the $\omega-\omega'$ plane}
Recently, many authors have investigated the evolution of
quintessence dark energy models in the $\omega-\omega'$ plane
\cite{18,19,20}. According to different regions in the
$\omega-\omega'$ plane, the models can be classified into two types
which are called thawing and freezing models \cite{17,18}.
Distinguishing the thawing class of dark energy from the freezing
class would reveal important clues to the nature of the new physics
\cite{18,19,20}.

In \cite{18}, R.R.Caldwell and E.V.Linder have analyzed the
following potentials for thawing behavior:
$V(\phi)=M^{4-\lambda}\phi^{\lambda}$, where $\lambda>0$ and
$V(\phi)=M^{4}\exp(-\beta\phi/M_{p})$. They have also analyzed the
following potentials for freezing behavior:
$V(\phi)=M^{4+\lambda}\phi^{-\lambda}$ and
$V(\phi)=M^{4+\lambda}\phi^{-\lambda}\exp(\alpha\phi^{2}/M_{p}^{2})$
for $\lambda>0$. They point out that if the result lies outside the
thawing and freezing regions then one may have to look beyond simple
explanations, perhaps to even more exotic physics such as a
modification of Einstein gravity. In what follows, we shall study
the cosmological dynamics of the NAQDE model in the $\omega-\omega'$
plane. For convenience, we will identify $\omega$ with $\omega_{q}$
and $\omega_{\phi}$.

At first, let us consider the expression of $\omega'$, where
$\omega'$ is the variation with respect to the e-folding time
$N\equiv\ln a$. From (12), one can obtain
\begin{equation}
\frac{d\omega}{dz}=\frac{2}{3n}\sqrt{\Omega_{\phi}}+\frac{1}{3n}\frac{1}{\sqrt{\Omega_{\phi}}}\frac{d\Omega_{\phi}}{dz}(1+z).\label{eq26}
\end{equation}

On the other hand,
\begin{equation}
\omega'\equiv\frac{d\omega}{dN}=-(1+z)\frac{d\omega}{dz}.\label{eq27}
\end{equation}

Substituting Eq.(\ref{eq26}) into Eq.(\ref{eq27}), we have
\begin{equation}
\omega'=-(1+z)[\frac{2}{3n}\sqrt{\Omega_{\phi}}+\frac{1}{3n}\frac{1}{\sqrt{\Omega_{\phi}}}\frac{d\Omega_{\phi}}{dz}(1+z)].\label{eq28}
\end{equation}

By employing Eqs.(\ref{eq10}), (\ref{eq12}) and (\ref{eq28}), we can
plot the trajectory of the NAQDE in $\omega - \omega'$ plane, as is
shown in Fig.4, where selected curves are plotted for the cases of
$n = 3.0, 3.5, 4.0$ and $5.0$ respectively. Here we take
$\Omega_{\phi0}=0.73$.

The blue line is the coasting line $\omega' = 3(1+\omega)^{2}$
following from constant field velocity, with $\omega'$ greater
(smaller) than this for field acceleration (deceleration). As shown
in Fig.4, we can see that the trajectory of the NAQDE lies below
coasting line at late times and the field accelerates.

The thawing region of phase space is defined by
\begin{equation}
1 + \omega \leq \omega' \leq 3(1 + \omega).\label{eq29}
\end{equation}

We note that the lower bound $1 + \omega$ is for $\Omega_{m0} < 0.8$
and $\omega < - 0.8$ at present.

Alternatively, the freezing region of phase space is defined by
\begin{equation}
3\omega(1 + \omega) \leq \omega' \leq 0.2\omega(1 +
\omega).\label{eq30}
\end{equation}

The upper bound $0.2\omega(1 + \omega)$ is for $\Omega_{\phi0} >
0.6$ and $\omega < - 0.8$ at present \cite{18,19}.

Next, we discuss the physical implication of the trajectory in
$\omega-\omega'$ plane.

At early times, as shown in Fig.4, the evolutionary trajectory in
the $\omega-\omega'$ plane lies outside the thawing and freezing
regions. We may ascribe this to the result of quantum corrections.
At late times, the trajectory of evolution enters the so-called
freezing region, gradually approaching to a static cosmological
constant state in the future. Roughly, the NAQDE should belong to
the freezing model at late times. Therefore, it turns out that the
quantum effect is more obvious at early times than at late times.

As we have seen, the NAQDE model in the $\omega-\omega'$ plane
clearly shows the evolutionary character of the dark energy, and the
dynamics of the NAQDE can be explored explicitly by reconstruction.

In addition, these results can be used to discriminate different
cosmological models and the future astronomical observations can
also discriminate the NAQDE models with different parameters
\cite{18,21}.

\section{the NADE with interaction}

Similar to \cite{4,21,22}, we further extend the NADE by including
the interaction between the NADE and a pressureless cold dark
matter, and investigate its evolution in $\omega-\omega'$ plane.

\subsection{The NADE with interaction}

We consider the interaction between NADE and DM such that $\rho_{m}$
and $\rho_{\phi}$ respectively satisfy
\begin{equation}
\dot{\rho}_{m}+3H\rho_{m}=Q,\label{eq31}
\end{equation}
\begin{equation}
\dot{\rho}_{\phi}+3H(1+\omega_{\phi})\rho_{\phi}=-Q,\label{eq32}
\end{equation}
where Q denotes the interaction term and can be taken as
$Q=3b^{2}H\rho$ with $b^{2}$ the coupling constant\cite{23,24}. A
more general consideration about the interaction term can be found
in\cite{25}.

In this case, by using Eqs.(\ref{eq7}), (\ref{eq8}), (\ref{eq5}) and
(\ref{eq31}), we find that the equation of motion for
$\Omega_{\phi}$ is given by
\begin{equation}
\frac{d\Omega_{\phi}}{da}=\frac{\Omega_{\phi}}{a}[(1-\Omega_{\phi})(3-\frac{2}{n}\frac{\sqrt{\Omega_{\phi}}}{a})-3b^{2}],\label{eq33}
\end{equation}

or
\begin{equation}
\frac{d\Omega_{\phi}}{dz}=-\Omega_{\phi}\{(1-\Omega_{\phi})[3(1+z)^{-1}-\frac{2}{n}\sqrt{\Omega_{\phi}}]-3b^{2}(1+z)^{-1}\},\label{eq34}
\end{equation}

From Eqs.(\ref{eq31}), (\ref{eq8}) and (\ref{eq5}), we obtain the
EoS as (we can also refer to \cite{4})
\begin{equation}
\omega=-1+\frac{2}{3n}\sqrt{\Omega_{\phi}}\frac{1}{a}-\frac{b^{2}}{\Omega_{\phi}},\label{eq35}
\end{equation}

or
\begin{equation}
\omega=-1+\frac{2}{3n}\sqrt{\Omega_{\phi}}(1+z)-\frac{b^{2}}{\Omega_{\phi}}.\label{eq36}
\end{equation}

As is referred in \cite{4}, if $b^{2}\neq 0$, from Eq.(\ref{eq35})
or (\ref{eq36}), one can see that $\omega$ can be smaller than $-1$
or larger than $-1$. As shown in Fig.$5$ and Fig.$6$, we can clearly
see that $\omega$ can cross the phantom divide.

Furthermore, as shown in Fig.$5$, we see that for a fixed
interaction parameter $b^{2}=0.10$ and $n=2.0, 2.5, 3.0$, $\omega$
is larger than $-1$ and crosses $-1$ earlier for larger n. But for
$n=4.0$ and $5.0$, we can see that $\omega$ is smaller than $-1$ at
early times. From Fig.$6$, for the constant $n=3.0$, except for
$b^{2}=0.10$, $\omega$ is smaller than $-1$ at early times and
crosses $-1$ then. We can also see that $\omega$ crosses $-1$ again
and earlier for larger interaction parameter $b^{2}$.

\begin{figure}
\center{
\includegraphics[scale=0.8]{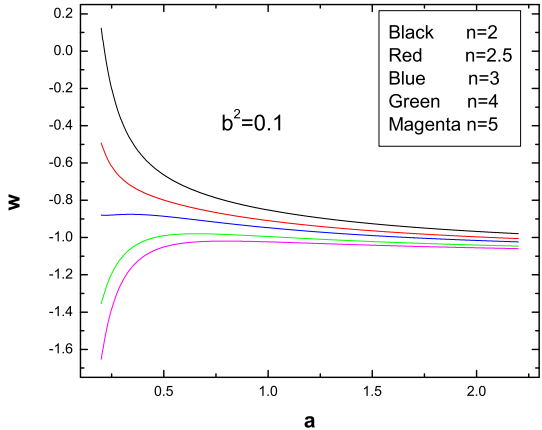}
\caption{\label{fig:wide}Behavior of the equation of state (35).
Here we take $\Omega_{q0} = 0.73$, and  show the cases for a fixed
interaction parameter $b^{2}=0.1$ but for different values of the
constant $n = 2.0, 2.5, 3.0, 4.0$ and $5.0$.} }
\end{figure}

\begin{figure}
\center{
\includegraphics[scale=0.8]{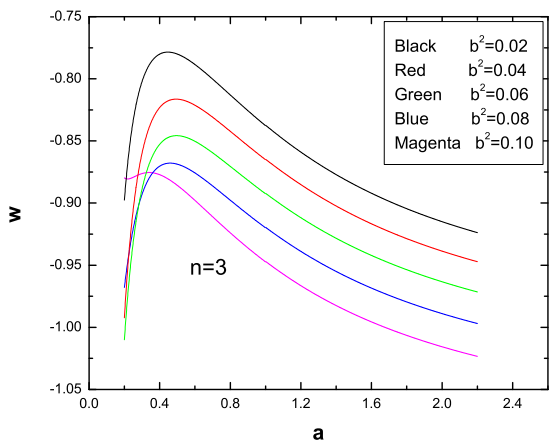}
\caption{\label{fig:wide}Behavior of the equation of state (35).
Here we take $\Omega_{q0} = 0.73$, and  show the cases for a fixed
constant $n = 3.0$ but for different values of the interaction
parameter $b^{2}=0.02, 0.04, 0.06, 0.08$ and $0.10$.} }
\end{figure}

\subsection{The evolution of the NADE with interaction in the $\omega-\omega'$ plane}

\begin{figure*}[tbph]
\center{
\includegraphics[scale=0.8]{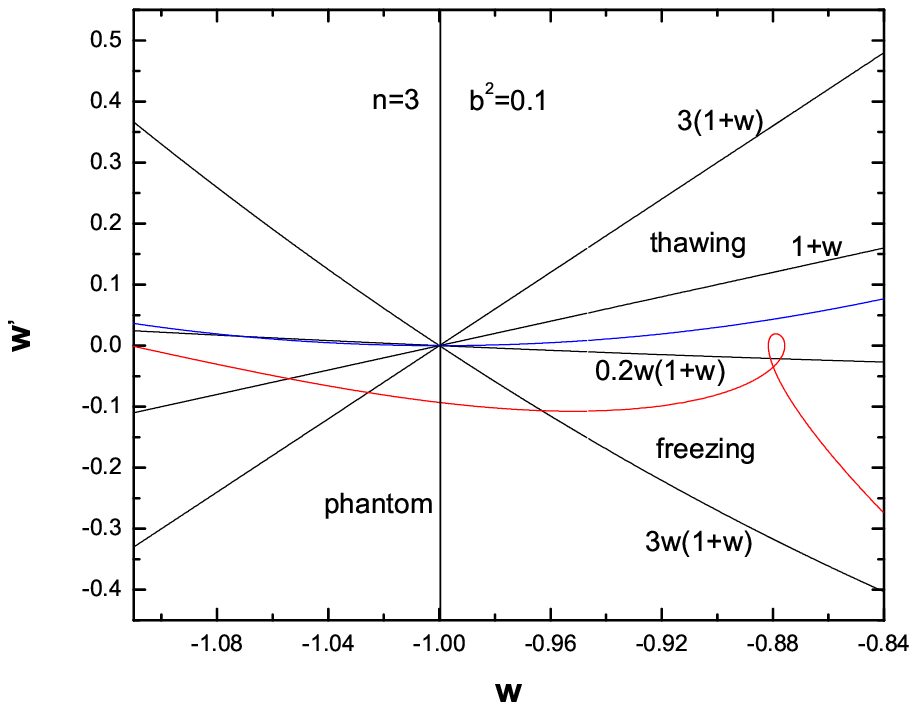}
\includegraphics[scale=0.8]{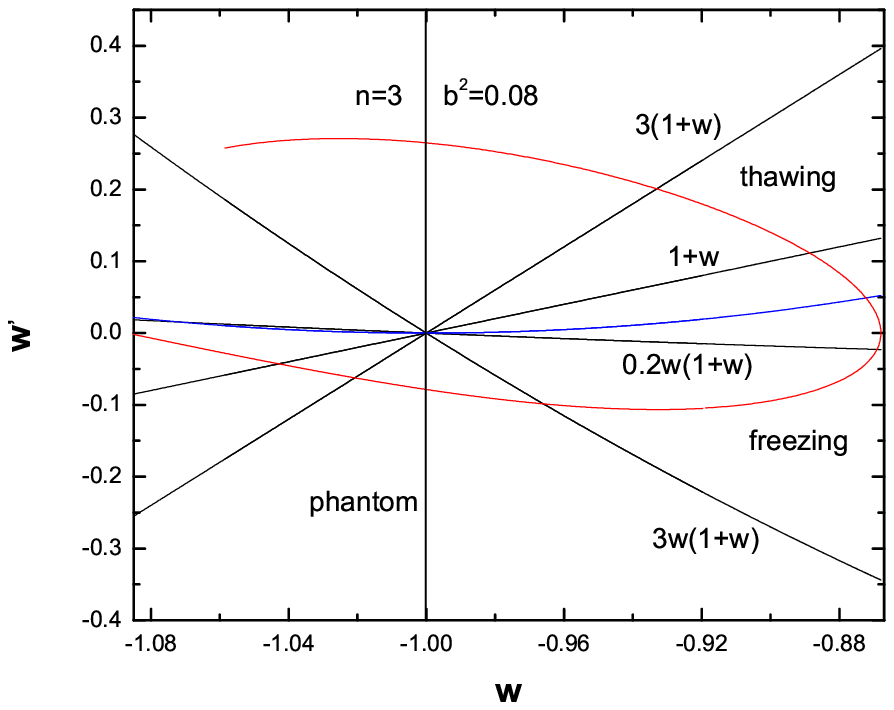}
\includegraphics[scale=0.8]{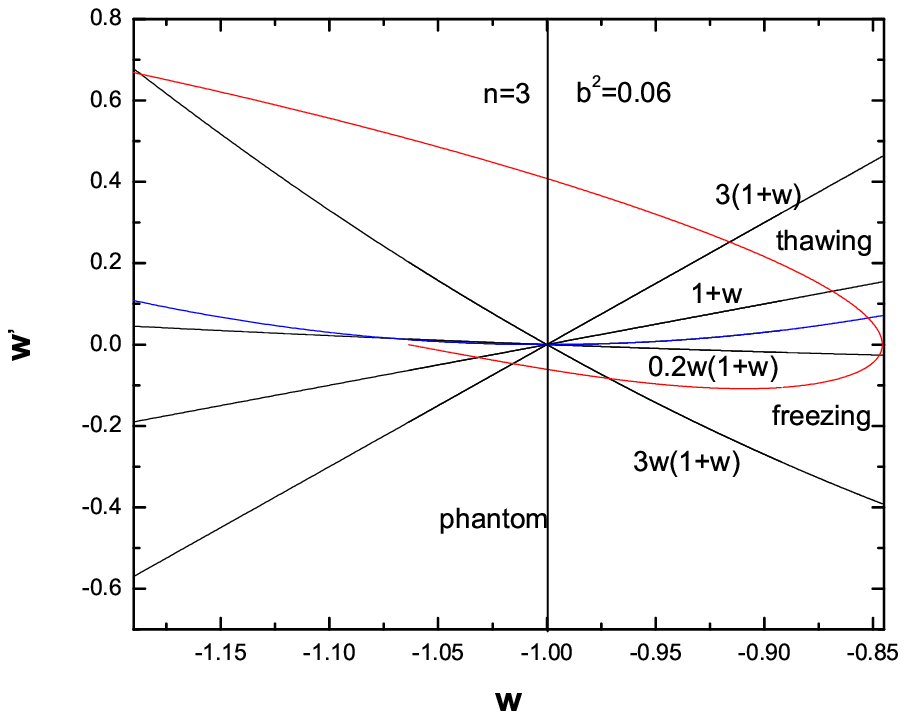}
\includegraphics[scale=0.8]{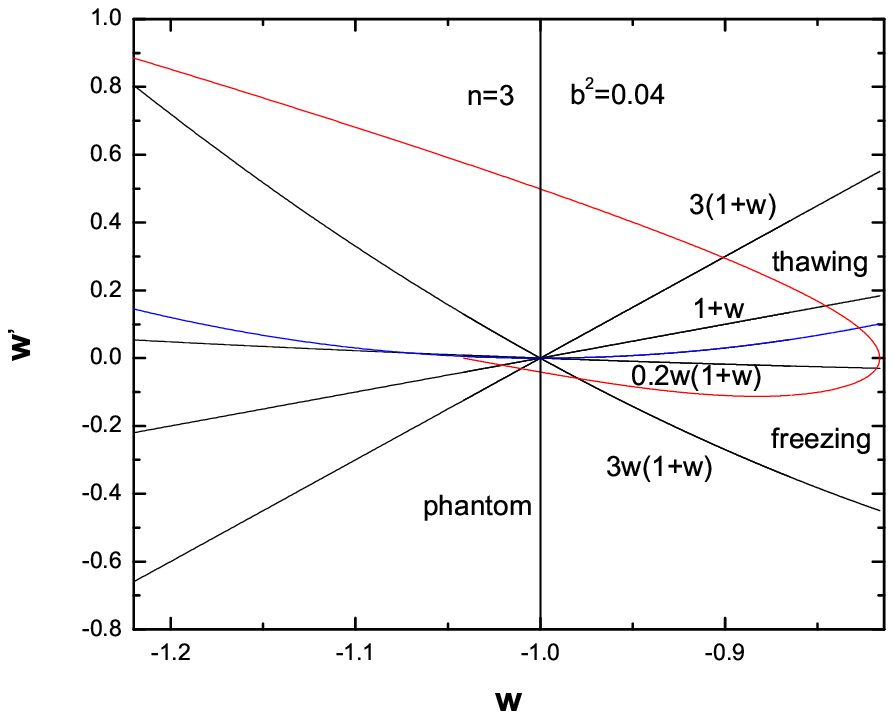}
\includegraphics[scale=0.8]{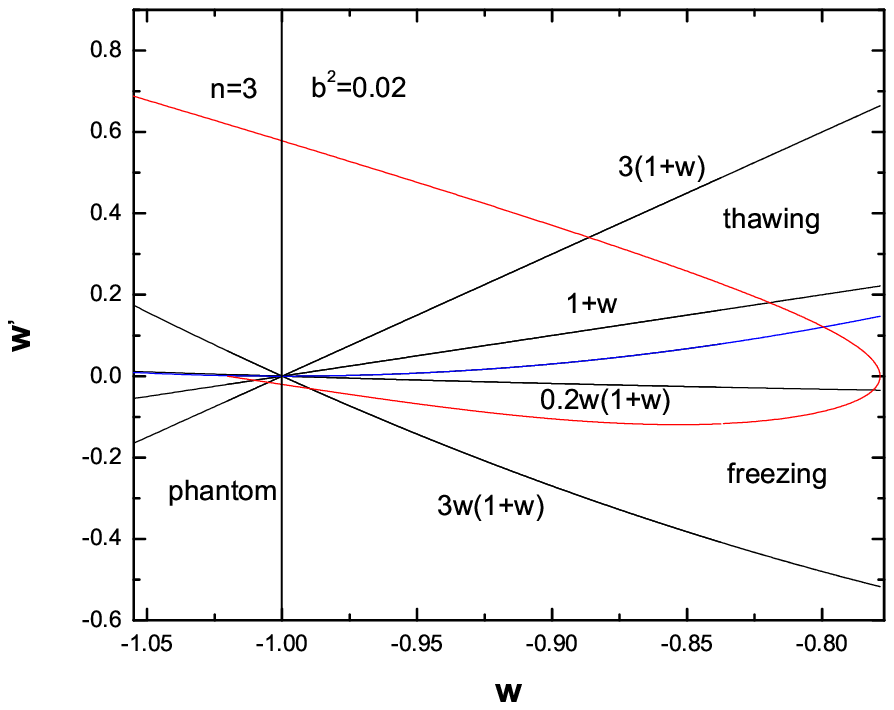}
\caption{The trajectory of the NADE with interaction in
$\omega-\omega'$ plane, where selected curves are plotted for the
cases of a fixed constant $n = 3.0$ but different values of the
interaction parameter $b^{2}=0.10, 0.08, 0.06, 0.04$ and $0.02$.
Here we take $\Omega_{\phi0}=0.73$.
 } }
\end{figure*}

In this subsection we investigate the evolution of the NADE with
interaction in the $\omega - \omega'$ plane.

At first, let us consider the expression of $\omega'$. From
Eq.(\ref{eq36}), one obtains
\begin{equation}
\omega'=-(1+z)[\frac{2}{3n}\sqrt{\Omega_{\phi}}+\frac{1}{3n}\frac{1}{\sqrt{\Omega_{\phi}}}\frac{d\Omega_{\phi}}{dz}(1+z)+\frac{b^{2}}{\Omega_{\phi}^{2}}\frac{d\Omega_{\phi}}{dz}].\label{eq37}
\end{equation}

By using Eqs.(\ref{eq34}), (\ref{eq36}) and (\ref{eq37}), we can
plot in Fig.$7$ the trajectory of the NADE with interaction in
$\omega-\omega'$ plane, where selected curves are plotted for the
cases of $n=3$ and $b^{2}=0.10, 0.08, 0.06, 0.04, 0.02$
respectively. Here we take $\Omega_{\phi0}=0.73$.

As shown in Fig.$7$, such coupling shift the trajectory of the NAQDE
in the $\omega-\omega'$ phase space, allowing for dynamics outside
the thawing and freezing regions. We can also see that except for
$b^{2}=0.10$, the trajectory lies above the coasting line at early
times and lies below the coasting line at late times. It elucidates
that the field acceleration at early times and deceleration in its
motion at late times. In the future, the trajectory crosses the
phantom divide and enters the phantom regions.

\section{Summary}
It is fair to claim that the NADE provides a reliable framework for
investigating the problem of dark energy, owning to possessing some
features of the quantum gravity theory. In this paper, we suggest a
correspondence between the new agegraphic dark energy scenario and
the quintessence scalar-field model. We have demonstrated that the
new agegraphic evolution of the universe can be described completely
by a single quintessence field. Its potential is reconstructed
numerically. In addition, we have also obtained the approximate
solution of the NAQDE in the matter-dominated epoch and analyzed its
character. Furthermore, we have investigated the evolution of the
NAQDE in the $\omega-\omega'$ phase plane. We find that the quantum
effect is more obvious at early times than at late times. For
comparison, the evolution of the NADE model with interaction has
also been investigated in the $\omega-\omega'$ plane.

This work is a first step towards studying the form of scalar
potential and the evolution in the $\omega-\omega'$ plane by quantum
corrections. It will also be interesting to further investigate the
analytical form of the NAQDE potential by some approximations. In
future works, we will extend such investigations including ADE, NADE
and holographic dark energy. In all this paper we assume our
universe is spatially flat but it is completely possible to show
that the parallel analysis could be extended to the spatially closed
and hyperbolic universe. We also expect that the further
investigation will provide us a more exact picture of dark energy.

\begin{acknowledgments}
We are grateful to Rong-gen Cai and Xin Zhang  for helpful
discussions, specially to Hao Wei for carefully reading the original
version of the manuscipt. This work is partly supported by
NSFC(Nos.10405027, 10663001), JiangXi SF(Nos. 0612036, 0612038) and
SRF for ROCS, SEM. We also acknowledge the support by the Program
for Innovative Research Team of Nanchang University.
\end{acknowledgments}

\newpage 
\bibliography{apssamp}

\end{document}